# Characterizing People's Daily Activity Patterns in the Urban Environment: A Mobility Network Approach with Geographic Context-Aware Twitter Data

Junjun Yin[1, *] and Guangqing Chi[1, 2]

1. *Social Science Research Institute and Population Research Institute*

2. *Department of Agricultural Economics, Sociology and Education*

*The Pennsylvania State University, University Park, PA, 16802, USA*

* To whom correspondence should be addressed. Email: jyin@psu.edu

**Abstract:** People's daily activities in the urban environment are complex and vary by individuals. Existing studies using mobile phone data revealed distinct and recurrent transitional activity patterns, known as mobility motifs, in people's daily lives. However, the limitation in using only a few inferred activity types hinders our ability to examine general patterns in detail. We proposed a mobility network approach with geographic context-aware Twitter data to investigate granular daily activity patterns in the urban environment. We first utilized publicly accessible geo-located tweets to track the movements of individuals in two major U.S. cities: Chicago and Greater Boston, where each recorded location is associated with its closest land use parcel to enrich its geographic context. A direct mobility network represents the daily location history of the selected active users, where the nodes are physical places with semantically labeled activity types, and the edges represent the transitions. Analyzing the isomorphic structure of the mobility networks uncovered 16 types of location-based motifs, which describe over 83% of the networks in both cities and are comparable to those from previous studies. With detailed and semantically labeled transitions between every two activities, we further dissected the general location-based motifs into activity-based motifs, where 16 common activity-based motifs describe more than 57% transitional behaviors in the daily activities in the two cities. The integration of geographic context from the synthesis of geo-located Twitter data with land use parcels enables us to reveal unique activity motifs that form the fundamental elements embedded in complex urban activities.

**Keywords:** *activity motif, activity pattern, geographic context, geo-located tweets, mobility network*

# 1. Introduction

Today's urban infrastructures and entities, such as residential places, working facilities, and shopping malls, scatter across the urban landscape to accommodate the needs of people's daily activities. Understanding detailed spatiotemporal activity patterns about how citizens interact with their surrounding environments is important for urban planning and its applications, such as characterizing urban spatial-temporal structures (Jiang, Ferreira, and González 2012b), assessing the impact of urban form on public health (Frank and Engelke 2001), and tackling sustainability issues of the urban systems (Ahas et al. 2010). Intuitively, people's daily activities in interacting with urban space vastly differ by individual. By using mobile phone call data for tracking people's movements, studies revealed that human movements in navigating urban space are highly predictable because of people's tendency to revisit places, which are known as preferential return behaviors (González, Hidalgo, and Barabási 2008; Song et al. 2010) and can be well modeled by a series of Markov chain-based models (Lu et al. 2013).

Recent studies continue to uncover unique and recurrent mobility patterns, known as the mobility motifs, in people's daily travel networks (Schneider, Belik, et al. 2013). For example, a typical home-to-work mobility motif represents the daily transitions between the home and work locations. That study suggested that 90% of people use just one of 17 mobility motifs in their daily lives, and nearly half of the population follows a simple two-location mobility motif.

Indeed, as humans are creatures of habit, distinct motif patterns are also observed in the transitions in people's daily activities by using travel survey data (Jiang, Ferreira, and González 2012a). Cao et al. (2019) suggested differentiating the two types of motifs as location-based motifs (LBM), which only concern the transitions among different locations, and activity-based motifs (ABM), which focus on the transitions among various types of activities.

One major research challenge found in the existing literature, as the premise to enable seeking spatiotemporal activity patterns, is the lack of information about detailed activity location and type. While such information is often collected in surveys, it is an expensive and labor-intensive process to survey a large population, and it often covers an individual's location history over only a few days. Despite privileged access to mobile phone data, the intrinsic low spatial accuracy on the order of several kilometers is often limited to inferring two major activity locations (i.e., home and work) and types (Jurdak et al. 2015; Huang and Wong 2016). Therefore, some studies began to explore publicly accessible location-based social media data with a higher spatial resolution, such as geo-located Twitter data and check-in records, as a means to track the locations of individuals. Many models and methods were developed to first derive the activity locations based on the agglomeration of recorded user locations and then infer the corresponding activity types (Liu et al. 2015; Sun et al. 2015; Jenkins et al. 2016). For example, a common approach is to identify the spatial clusters of the record locations to infer the activity locations in people's daily lives. Owing to the lack of geographic context associated with recorded locations as semantic labels, it is difficult to differentiate one spatial cluster from another, let alone to tie specific activity types to the derived activity locations. However, recent studies have revealed unique spatiotemporal signatures tying collective human activity patterns to different land uses in the urban environment (Liu et al. 2015; Soliman et al. 2017). Studies on human activity patterns also benefit from integrating land use information as geographic context to infer activity types (Soliman et al. 2015; Huang and Wong 2016). Ideally, tracking the movement of individuals with geographic context-aware user locations will bring more clarity to each transition between two activities and enable us to examine the activity transition patterns (e.g., ABMs) in greater detail.

In this study, we propose a mobility network approach to studying people's daily activity patterns in the urban environment. More specifically, we aim to discover activity patterns concerning both LBMs and ABMs for characterizing the transitions in people's daily activities. To enable this study, we utilized publicly accessible geo-located Twitter data with a high spatial accuracy to track the movements of individuals in the urban environment. To enrich its geographic context, we linked each recorded user location with its closest land use parcel. Eleven activity types (e.g., home, work, and shopping) were defined based on their correspondence to the land use information, which was used to construct the geographic context-aware daily location history for each Twitter user.

To address daily activity patterns, we defined the criteria for selecting active users whose daily whereabouts could be identified. The details of such criteria are introduced in Section 3.2. By analyzing the shape of the collective location histories of the active users, we identified preferential return patterns and confirmed the existence of recurrent transitional behaviors. A graph was then constructed to represent each user's daily location history as a mobility network. For LBMs, the nodes were the visited activity locations, and the edges were the movements transitioning from one location to another. In contrast, for ABMs, the nodes were semantically labeled activity types, whereas the edges represented the transitions among those activities. Further exploration into the isomorphic structure within the daily mobility networks uncovered detailed LBMs. With the transitions labeled by detailed activity types, we further dissected general activity patterns into detailed ABMs for characterizing transitions among different activities in people's daily lives. The study provided us with deeper insights into the unique activity motifs that form the fundamental elements embedded in complex urban activities. The case studies conducted in two U.S. cities Chicago and Greater Boston suggested that our findings with geo-located Twitter data were consistent across geographic space.

The remainder of this manuscript is organized as follows. Section 2 describes the background and related work about the activity patterns of individuals in urban environments. Section 3 details the data and methods used in this study for capturing and measuring the spatiotemporal activity patterns of individuals in the urban environment. Section 4 presents a detailed analysis of the spatial and temporal activity patterns of the individuals in Chicago and Greater Boston. Section 5 concludes this study with discussions about the contributions and limitations of our mobility network approach.

# 2. Background and Related Work

## Spatiotemporal human activity patterns and activity motifs

Human activities are complex behaviors and are confounded by human variability. To seek spatiotemporal human activity patterns in the urban environment, existing research primarily encompasses two main perspectives: the collective level and the individual level (Chi and Zhu 2019). Collectively, various types of human activities take place across the urban space simultaneously, such as people staying at home, taking trips at transportation sites, and shopping at malls. Researchers attempt to understand the relationships between the agglomeration of people's activity locations and the configurations of the urban environment (Liu et al. 2015).

Individually, people navigate through those activities at different locations based on their own time schedules and routines. The recent research efforts focus on seeking common structures that can be used to characterize the transition patterns of individuals in moving across geographic spaces. Determining whether the spatial configuration of the urban environment confines human activities in it or whether human activities shape and define the image a city is an ongoing philosophical and scientific quest in understanding how cities function (Lynch 1960). Nevertheless, because human activities in the urban environment are essentially different forms of spatial interactions between citizens and the urban infrastructures and entities, insights gained from both collective and individual perspectives can complement each other to provide a more holistic view.

Abundant emerging studies have uncovered new insights into individual human activity patterns in the urban environment, using various types of data sources and methods. For example, GPS (Global Positioning System) tracks of residents were used to study daily mobility patterns between two local communities (Davies et al. 2019), longitudinal survey data were used to explore the community structures in people's collective activity space (Xi, Calder, and Browning 2020). González, Hidalgo, and Barabási (2008), using mobile phone call data in the form of call detail records (CDR), revealed that trajectories of human movements across the urban space are not random but show a high degree of spatial and temporal regularity. The mobility patterns of individuals characterized by time-independent travel distances revealed preferential return behaviors, suggesting that people are likely to revisit a few highly frequented locations. The authors analyzed the shapes of the trajectories of individuals and found out that people's movements tend to center around the principal axes determined by these dominated locations. The inherent anisotropy of each trajectory when aligning them in the same intrinsic reference frame could be described by a single spatial probability distribution, which indicated that humans follow simple reproducible patterns despite their different travel histories. Further investigations into preferential return behaviors, using mobile phone data, confirmed the existence of recurrent transitions among those frequently visited locations, which suggested that people's spatial and temporal visitation behaviors are highly predictable (Song et al. 2010) and can be well modeled by a series of Markov chain-based models (Lu et al. 2013). Similar mobility patterns at a larger spatial scale were observed by using geo-located Twitter data (Jurdak et al. 2015). Although these studies focused on people's movements without explicitly examining the exact activity types associated with the movements, the developed methods and methodologies laid the foundations to advance recent studies on human activity patterns.

Therefore, acquiring information about the specific activity types at the visited locations becomes a necessary step before activity pattern analysis. Such information is usually collected by explicitly asking in surveys and questionnaires, or it is inferred using a set of heuristics tailored to datasets used to track the movements of individuals. For example, the most frequent mobile phone call location at night indicates that the individual is at home, while the second most frequent location is the workplace (Ahas et al. 2010; Kung et al. 2014). When geo-located social media data are used, the home and work locations are often inferred by the locations with the most and the second-most social media posts, such as when people use check-ins (Cho, Myers, and Leskovec 2011) and geo-located Twitter data (Luo et al. 2016). Using travel survey data to analyze activity-annotated location histories of individuals allows identification of an activity-based signature of daily travel patterns (Jiang, Ferreira, and González 2012a).

Further, by modeling the transitions among individuals' activity locations as travel networks, researchers found unique daily urban mobility motifs, which were evident from a week-long survey dataset (Schneider, Rudloff, et al. 2013) and mobile phone call data (Schneider, Belik, et al. 2013). The notion of motif is derived from the term "network motif", which is defined as the recurrent and statistically significant subgraph within a network collection (Newman 2018). For example, gene expression motifs were extracted from gene expression data that a subset of genes can be simultaneously conserved across many samples (Chechik et al. 2008). Motifs were also used as the overrepresented DNA sequences to identify gene expression patterns (Lu et al. 2013). Although people's daily activities in the urban environment are complex behaviors, the prominent motifs in people's daily travel networks transitioning from one frequently visited activity location to another indicate that their daily activity patterns are highly generalizable and can be depicted by a few motifs. For example, Schneider, Belik, et al. (2013) suggested that 90% of people use one of just 17 mobility motifs in their daily life. Furthermore, motifs in human daily activities have been observed in multiple cities. Jiang, Ferreira, and González (2017) developed a framework to derive mobility motifs using mobile phone CDR data in Singapore and suggested that the results were comparable to the ones from travel survey data. However, the mobility motifs extracted in those studies did not identify the specifications of the activity at the locations. For example, a two-node mobility motif can be a "home-to-work" transition but can also be a "home-to-school" transition. Indeed, these mobility motifs were conceptualized as LBMs, where the nodes of the motifs are frequently visited activity locations but with unknown activity types. In contrast, the nodes in ABMs are the locations with annotated activity types (Cao et al. 2019).

To study ABMs, it is logical to first get a better sense of the visited locations and then infer the corresponding activity types, which is in line with the research interests in seeking human activity patterns at the collective level. At that level, the research questions often focus on the linkages between the agglomerations of human activity space and the characteristics (e.g., functionality and structure) of urban areas. The fundamental idea is that people's activities are tightly connected with the socio-economic features of the environment and therefore reflect the spatial configurations of the physical space (Liu et al. 2015). For example, Jiang et al. (2015) developed a method for urban land use classifications by using online point-of-interest data for inferring people's activity locations. Mobile phone call data were used to infer human activities for revealing the spatiotemporal structure of urban areas (Jiang, Ferreira, and González 2012b), geo-located tweets were collected to examine the alignment between Twitter hotspots and corresponding physical locations across multiple cities (Jenkins et al. 2016), check-in data were used to generate the movement flows for identifying city centers (Sun et al. 2015), and geo-Weibo (micro-blogs with geo-tags) data were used to extract and analyze a city's tourism districts (Shao, Zhang, and Li 2017).

Examining the aggregated/spatially clustered human activities provides a bird's-eye view in terms of the correspondence of those activities to urban structures/functionalities. In turn, the land use information of the urban environment can be integrated as a geographic context to each activity location to better understand the activity patterns of individuals. For example, Huang and Wong (2016) used an urban land planning map to associate the locations of Twitter users with one of the four activity zones to characterize their activity space. In another study, Soliman et al. (2015) used parcel-level detailed land use maps to understand activity types related to the locations of Twitter users in Chicago. Although the main scope of this study is on activity patterns at the individual level, we aim to connect the research in studying collective human activity patterns to better characterize people's daily activity patterns in the urban environment.

# Geo-located Twitter data for studying human activity patterns

Studies on detailed human activity patterns in the urban environment rely on data sources that are capable of tracking individuals moving across the urban space (Huang and Wong 2016). Conventional data sources are available in the form of travel surveys (Jiang, Ferreira, and González 2012a) and activity diary records (Chen et al. 2011). However, it is labor-intensive and expensive to survey a large group of people or monitor their movements for a relatively long period. The existing literature suggests that mobile phone call data are a popular data source for examining detailed human activity patterns across a larger spatial scale. However, high-quality mobile phone data are privileged information, and researchers have only limited access because of privacy concerns (Crawford and Finn 2015; Huang and Wong 2016). The low spatial accuracy of mobile phone data, on the order of several kilometers (Jurdak et al. 2015), also makes it difficult to infer activity types of recorded locations.

Consequently, many recent studies utilize publicly accessible location-based social media data with a higher spatial accuracy, such as geo-located tweets and check-in records, to study human activity patterns. In particular, geo-located Twitter data have been proven to be a useful data source for studying human mobility patterns at large spatial scales (Hawelka et al. 2014; Yin et al. 2016), and the findings are comparable to those of studies that used CDR data (Jurdak et al. 2015). Check-in records have enabled further analysis of the characteristics of human activity patterns: they provide information about the venues where a check-in was made, such as whether it is a residence, an arts and entertainment center, or a business venue. To some extent, venue information serves as the geographic context of user location, which can be linked to the particular activity a user is engaging in at that moment. As previously mentioned, check-in data are popular for reflecting urban activity patterns at the collective level, but because of data sparsity, they may not be suitable for studying mobility and activity patterns at the individual level. For example, users post check-in messages when visiting a place of interest (usually a place that is new to them). It is unlikely, though, that anyone would frequently check in at a previously visited place and/or at a home or workplace. Nevertheless, it inspires a logical step for enhancing geographic context in geo-located tweets by developing a similar approach employed in the check-in data.

## 3. Materials and Methods

### Geo-located tweets and parcel-level land use maps

This study used geo-located tweets for characterizing people's daily activity patterns in Chicago and Greater Boston from January 1 to December 30, 2014. The data collection utilized the Twitter Streaming API (https://developer.twitter.com) by setting up two geographical bounding boxes and retrieving all geo-located tweets that fell within them. The bounding boxes used the lower-left and upper-right coordinates for Chicago (41.20, –88.70; 42.49, –87.52) and Greater Boston (41.41, –72.66; 43.12, –69.45). Because data collection was done at the city level, it did not exceed the data volume limit (i.e., 1% of the total real-time tweets generated on twitter.com) mentioned in Hawelka et al. (2014). In other words, we were able to collect almost all available geo-located tweets over the two cities. The entire data collection contains over 10.2 million geo-located tweets from Chicago and 12.5 million from Greater Boston.

The original location information embedded in the geo-tag is given in units of latitude and longitude. We examined the "geo" attribute in each raw tweet and kept the one with location information derived from the GPS receiver rather than from the geocoding process. Note that many tweets are generated by non-human Twitter users (i.e., bots). In the case of geo-located tweets, the geo-tagged locations are either stationary (tagged with the same geo-location) or have excessive relocating speed (one tweet with the same message content but with multiple preset geo-locations for broadcasting purposes). To exclude non-human Twitter users, we filtered the raw tweets in five steps. First, we removed duplicated tweets and kept those that fell within the administrative boundaries of Chicago and Greater Boston. Second, we filtered out tweets related to advertising (e.g., job, recruiting, and hiring) and broadcasting activities (e.g., weather and traffic warnings) because those tweets were often generated by artificial Twitter accounts (Hawelka et al. 2014). Third, for each remaining tweet, we performed a geospatial operation to search for its nearest land use parcel, and we assigned the land use type to the tweet location. A detailed description of this process is provided in the next section. We constructed a location history for each Twitter user by appending all the recorded locations in chronological order, sorted by timestamps. To remove artificial Twitter accounts with stationary locations, we identified the accounts with the same locations labeled with a land use type other than residential, which was one of the benefits of using land use parcels to infer the geographic context of tweet locations. Fourth, we removed non-human users based on unusual relocation speed by examining consecutive locations of each user and excluding those with relocation speed over the threshold of 240 m/s used by Jurdak et al. (2015). Finally, to reflect the activity patterns of residents rather than tourists, we imposed a criterion that a user who has stayed in the study region more than 30 days is considered as a citizen, as suggested by Yin et al. (2017). Note that this 30-day criterion was subjectively defined but strict to ensure that the observed Twitter population is actively observed in the study regions. At that stage, our data contained 87,866 and 98,024 individual Twitter users from Chicago and Greater Boston, respectively. The spatial coverage of the filtered geo-located tweets from the two cities is shown in Figure 1, where each point corresponds to a geo-located tweet collected for this study.

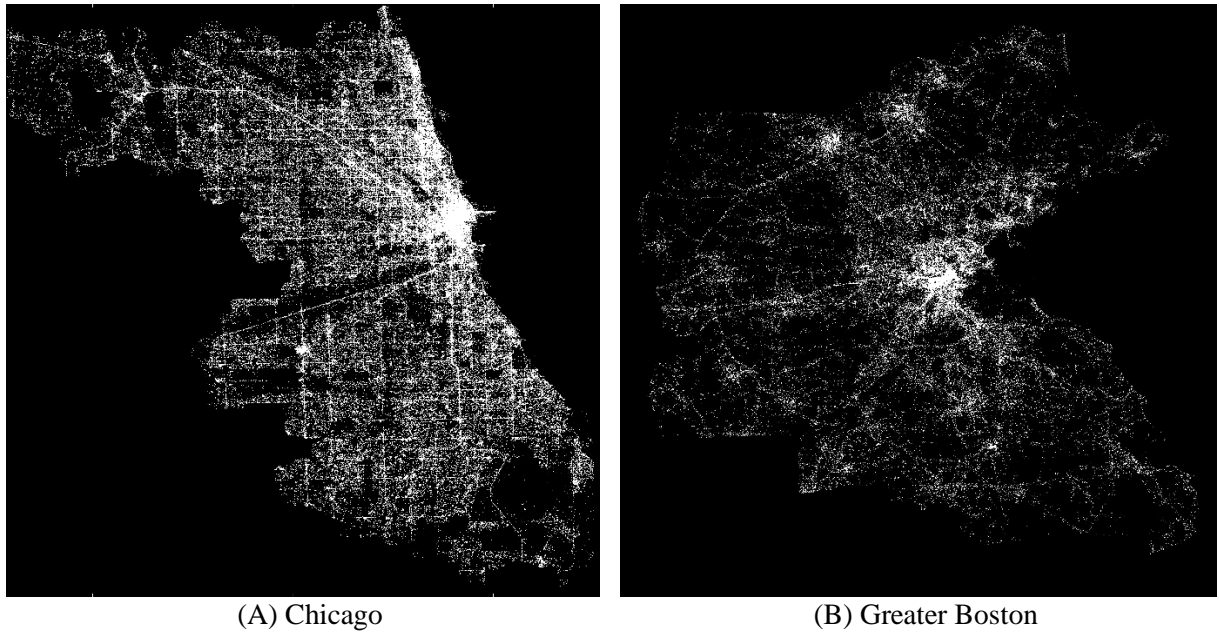

(A) Chicago          (B) Greater Boston

**Figure 1:** Spatial coverage of the geo-located Twitter data across the two cities in this study (A) Chicago and (B) Greater Boston

The reasons for selecting Chicago and Greater Boston as the study regions are twofold. First, previous studies seeking human activity patterns (e.g., LBMs) were conducted in Chicago using travel survey data (Jiang, Ferreira, and González 2012a; Schneider, Rudloff, et al. 2013) and in Boston using mobile phone data (Jiang et al. 2016). It sets the foundation that the results from our study using geo-located Twitter data may be comparable to the ones from using mobile phone and/or travel survey data. Second, this study employed parcel-level detailed land use maps to infer the geographic context of tweet locations, but the complex land uses in densely built metropolitan areas, such as Manhattan in New York City, could induce significant uncertainty. For example, a tall building in a downtown area can have different uses from the ground floor and up, which is often labeled "urban mix" for the land use category. Note that while the situation is relatively less complicated in Chicago and Greater Boston, those land use parcels labeled as "urban mix" remained in this study.

Urban land use in any city is expected to be constantly changing. The land use maps in our study should have been updated as close as possible to 2014 (i.e., the same year of our Twitter data collection). Parcel-level detailed land use maps of Chicago were not available for 2014 but had been updated through 2013 and were extracted from the Land Use Inventory for Northeast Illinois, 2013 (CMAP 2019). Land use maps of 2014 covering Greater Boston were extracted from the MassGIS (Bureau of Geographic Information) Level 3 Assessors' Parcel Mapping dataset (MassGIS 2019). For Chicago and Greater Boston, there are 164,619 and 1,116,482 polygonal land use parcels, respectively. Each dataset is accompanied by a coding scheme that defines the corresponding land use categories (e.g., there are 60 land use classes in the Chicago dataset).

To reflect the activities in those land use parcels, we adopted the activity scheme developed by Jiang, Ferreira, and González (2012a) for inferring activity types from travel surveys, which consisted of nine aggregated activity classes. Further, we made some modifications to the activity scheme in reclassifying land use to activity types as suggested by Soliman et al. (2017). In specific, we separated school activity into K-12 schools and universities/colleges because those two activities can differ vastly, and the detailed land use parcels were able to help us identify which was which. Also, we added the label "mixed-use" to urban-mix land use parcels. Finally, because hotel/resort activities were missing in travel surveys, we listed them as an individual activity class. The relationships between the activity code and land use category and the percentage of each type of land use parcels are shown in

Table 1. Notice that the residential land use parcels are the most prominent urban features in both datasets.

**Table 1:** Activity code and percentage of the land use parcel categories in Chicago and Greater Boston

| Activity Code | Land Use Category | Chicago | Greater Boston |
|---|---|---|---|
| 1 | Residential | 74.15% | 92.65% |
| 2 | Hotel/Resort | 0.12% | 0.04% |
| 3 | Mixed-Use | 12.36% | 0.80% |
| 4 | K-12 Schools | 0.79% | 0.10% |
| 5 | University/College | 0.15% | 0.11% |
| 6 | Office/Workplace | 2.71% | 1.36% |
| 7 | Services | 0.50% | 0.56% |
| 8 | Civic/Religious | 1.91% | 0.26% |
| 9 | Shopping/Retail | 0.07% | 0.99% |
| 10 | Recreation/Entertainment | 0.85% | 0.66% |
| 11 | Transportation | 3.49% | 0.55% |
| 12 | Others | 2.90% | 1.94% |
| Total number of parcels | | 164,619 | 1,117,027 |

## Activity locations and active Twitter users

For each location in the constructed trajectory, we performed a geospatial operation searching for its nearest land use parcel within 250 meters of its radius. The distance value of 250 meters was set to account for the inaccuracy of locations read from the GPS units in mobile devices (Jurdak et al. 2015). If there was no parcel within the radius, the activity code assigned to the location was set to 12 as unknown (i.e., "others"). Each trajectory was transformed as the activity sequence of an individual, where each activity was inferred from its corresponding geographic context-aware location. It is important to note that a Twitter user posting a tweet at those locations does not necessarily mean those are the locations at which normal daily life activities take place (this concern also applies to mobile phone call locations). Although random locations do exist, many studies have shown that frequently visited locations provide prominent correspondence to users' activity locations, owing to people's tendency of returning to previously visited places. Previous studies relied on spatial clustering methods to derive users' active locations, such as the density-based spatial clustering of applications with noise method (Jurdak et al. 2015; Shao, Zhang, and Li 2017; Soliman et al. 2017). However, those methods treat each location equally and often rely on a global parameter in determining the clusters, which is problematic between dense-built and less dense-built urban environments (Kwan 2016). In this study, because each location in a user's trajectory was anchored to the nearest land use parcel, two criteria were applied to active locations: (1) only those land use parcels with tweet counts above the average number of tweets in a user's trajectory were considered active locations; and (2) the popularity of those active locations were ranked by the number of tweets at each parcel. This approach serves as a de facto spatial clustering method that also benefited the inference of a user's home location. Because the most tweeted locations are not necessarily users' home locations, the home locations were determined based on the following rules. First, only the residential land use parcel with the most tweets between 9:00 pm and 6:00 am was chosen as the user's home location. Second, if the first condition was not met, the residential land use parcel with the highest rank was chosen as the user's home location. Otherwise, a user's home location was set to be unknown and excluded from this study.

     Before we analyzed activity patterns, we made note of the representativeness issues in the Twitter user population versus the actual population. Although addressing the representativeness of the Twitter user population is outside the scope of this study, we performed a regression analysis of the Twitter user population versus the actual population at the census tract level for the two cities. The Twitter user population in a census tract is the number of Twitter users with

their home locations falling within the census tract and the actual population was obtained from the American Community Survey (ACS) 2012-2016 five-year estimates (U.S. Census Bureau 2016). Figure 2 showed that the Twitter user population in this study correlated well with the ACS population estimates at the census tract level, which is measured by Pearson's *r* value 0.72 in Chicago ($p < 0.001$) and 0.71 in Greater Boston ($p < 0.001$). Although this approach does not fully support generalizing the findings in this study to the whole population, it does indicate the extent of the Twitter user population representing the whole population of the two cities in this study.

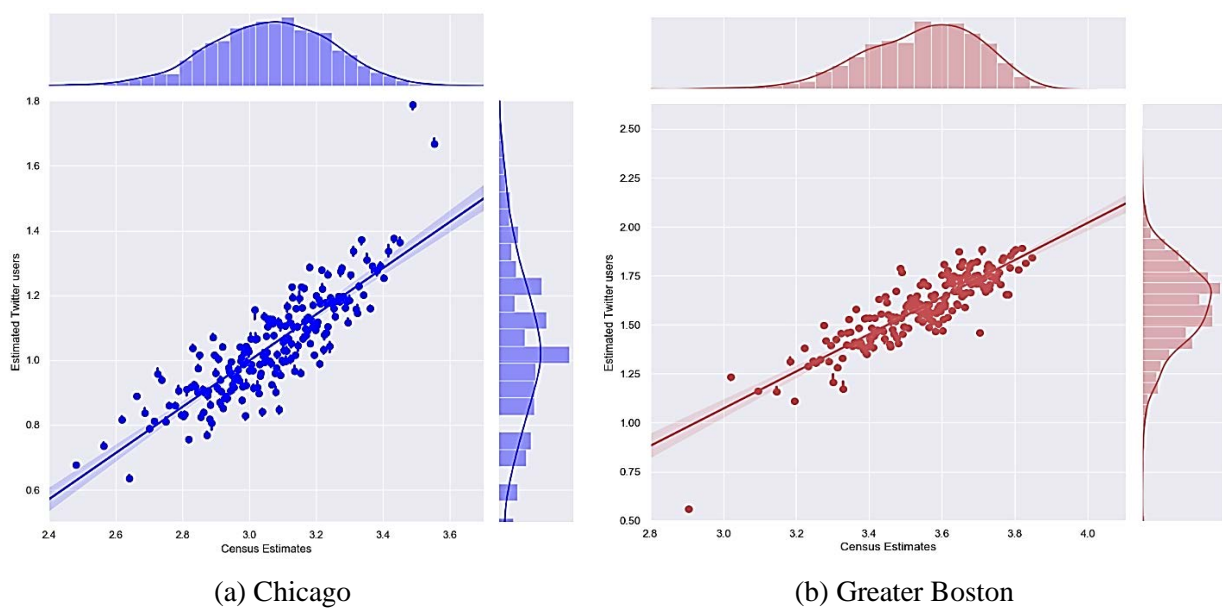

(a) Chicago  (b) Greater Boston

**Figure 2:** Representativeness of the Twitter user population versus actual population at the census tract level in (a) Chicago (Pearson's r = 0.72) and (b) Greater Boston (Pearson'' r = 0.71). The x-axis is the population estimates from the census, and the y-axis is the estimated number of Twitter users

One critical data processing step in this study was to select active users. As mentioned above, when digital geo-located data (e.g., CDR data and geo-located tweets) were used as proxies for tracking people's locations, the latent locations between two observed locations can be missing. To study people's daily activity patterns, it is necessary to impose a stricter criterion to minimize the impact of missing latent locations. Indeed, when addressing such an issue with mobile phone data, researchers conclude that a user can be considered actively observed (i.e., an active user) when she/he is found in at least 8 out of 48 time slots during an entire data observation (i.e., 24 hours into 48 half-hour time slots) (Schneider, Belik, et al. 2013; Jiang, Ferreira, and González 2017). Considering the location information is only available when people tweet, using at least 8 tweets per day as a selection criterion will make the Twitter user population skew away from the general population. Therefore, we relaxed the criterion and considered a user as active (with her/his locations and transitions actively observed) only when observed in at least 6 of the daily 48 time slots.

## Mobility networks and activity motifs

In this study, we constructed a mobility network for every individual active user's daily location history. Each mobility network was represented by a directed graph $G \equiv \langle V, E \rangle$, where $V$ is a set of nodes corresponding to the activity locations, and $E$ is a set of edges representing the transitions between the node pairs. Note that the property of node $V$ regarding the activity type associated with the location was ignored in searching for LBMs, but it was explicitly considered in searching for ABMs. From the perspective of seeking activity patterns, the tasks can be naturally translated into finding common structural patterns among a collection of mobility networks. In the cases of pursuing activity motifs, it is essentially a process of finding the recurrent subgraphs (also known as network motifs) to represent the common transition sequences in people's daily activities. For example, let $\{G_1 \equiv \langle V, E \rangle, G_2 \equiv \langle V, E \rangle, G_i \equiv \langle V, E \rangle, i = 1, 2, \ldots n\}$ be a mobility network collection representing a person's daily location histories in $n$ days or $n$ persons' location histories in one day. The activity motifs to be sought (i.e., LBMs and ABMs) should correspond to the network motifs among the network collection. To better explain the concept of the mobility network and the differences between LBMs and ABMs, two scenarios of daily activities are illustrated in Figure 3. The first scenario is a two-node activity motif, whereas the second scenario is a four-node activity motif. The nodes in the mobility networks are physical locations (i.e., places), and the edges are the transitions representing people's visits among those locations. The difference between an LBM and an ABM is that the node types are explicitly annotated in the case ABMs.

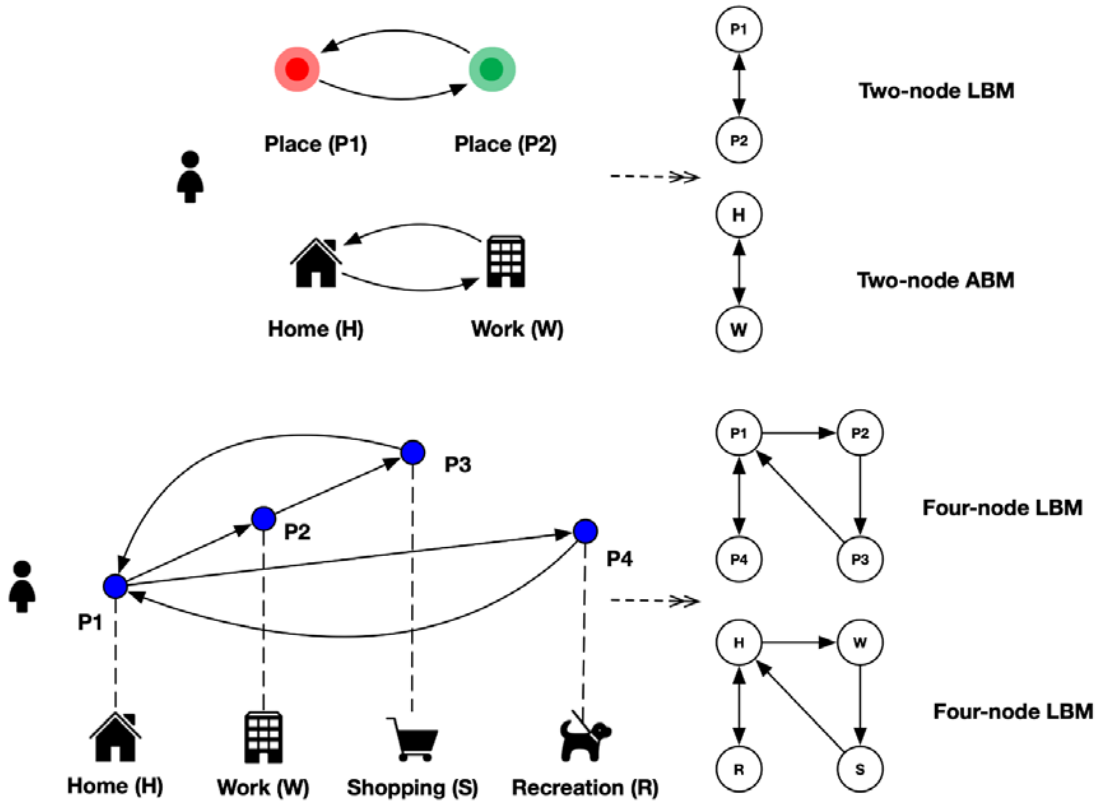

**Figure 3**: Illustration of peoples' daily activity motifs (LBMs and ABMs)

In graph theory, network motifs can be identified by exploring the network isomorphism (Newman 2018). Let $G_1 \equiv \langle V^1, E^1 \rangle$ and $G_2 \equiv \langle V^2, E^2 \rangle$ be two graphs. If $V^2$ is a subset of $V^1$ (noted as $V^2 \subseteq V^1$) and $E^2$ is a subset of $E^1$ ($E^2 \subseteq E^1$), then $G_2$ is a subgraph of $G_1$. Note that there are two types of subgraphs: vertex induced (or simply induced) and edge induced. A vertex-induced subgraph is a subset of the vertices of a graph $G$ together with any edges whose endpoints are both in that subset, whereas an edge-induced subgraph is a subset of the edges of a graph $G$ together with any vertices that are their endpoints. In this study, we considered only vertex-induced subgraphs because both LBMs and ABMs are vertex-based (i.e., node-based) sequences. If $G_2$ is a vertex-induced subgraph of $G_1$, and there is a one-to-one (i.e., bijection) mapping function $f: V(G_2) \rightarrow V(G_1)$ in which any two vertices $u$ and $v$ of $G_2$ are adjacent if and only if $f(u)$ and $f(v)$ are adjacent in $G_1$, then $G_1$ and $G_2$ are considered isomorphic ($G_2 \leftrightarrow G_1$). The mapping $f$ is called an isomorphism between $G_1$ and $G_2$. When there is a subgraph $G_1'$ of $G_1$ ($G_1' \subset G_1$) and $G_1'$ is isomorphic to $G_2$, it means an appearance of $G_2$ in $G_1$. The total number of appearances is the frequency $F_G$ of $G_2$ in $G_1$. Once the frequency $F_G(G_2)$ exceeds a predefined cut-off value, $G_2$ is considered a recurrent/frequent graph in $G_1$. Such a cut-off value is often defined as the arithmetic mean frequency $F_G(G_2)$ in a set of $N$ randomized graphs generated from $G_1$ using the null model (Milo 2002). Given the fact that randomized mobility networks are not meaningful, motifs in this study were the daily mobility networks whose frequencies were more than 0.5 percent of the total number of mobility networks in the dataset (Schneider, Belik, et al. 2013).

There are various existing implementations of network motif discovery algorithms (Masoudi-Nejad, Schreiber, and Kashani 2012). In particular, some recent network motif discovery algorithms were developed for mobility networks (Schneider, Belik, et al. 2013) and temporal networks (Paranjape, Benson, and Leskovec 2017). While these algorithms could potentially be employed in this study for analyzing LBMs, it is difficult to customize the inherent data structure to include an extra attribute of the nodes (i.e., activity type) in the case of ABMs. Therefore, this study utilized the improved graph-matching algorithm, named VF2 (Cordella et al. 2001), to detect the network motifs, which was implemented in NetworkX (Hagberg, Schult, and Swart 2008). Note that the detection of network motifs is computationally challenging. Because network motifs are considered the fundamental elements that uncover the structural design principles of the network collection, they are often subgraphs with a limited number of nodes (Milo 2002). For example, the maximum number of nodes in the activity motif detected using mobile phone data was set at 6 by Schneider, Belik, et al. (2013) and at 5 by Jiang, Ferreira, and González (2017). Therefore, we set the maximum number of nodes in a motif at 6.

## 4. Results

### Preferential return behaviors

In this study, mobility networks are used to represent the transitions in people's daily activities in the urban environment. By definition, activity motifs represent recurrent transitions among frequently visited locations/activities. In other words, the existence of activity motifs depends on the premise that people are likely to revisit a few frequently visited locations (i.e., preferential return behaviors). Previous work has demonstrated that confirmation of preferential return behaviors can be achieved by analyzing the shapes of individuals' trajectories/location histories (González, Hidalgo, and Barabási 2008). Suppose an active Twitter user's trajectory (i.e., location history) is a set of $(x, y)$ pairs $\{(x_1, y_1), (x_2, y_2), \ldots, (x_n, y_n)\}$, where $(\bar{x}, \bar{y})$ is the center of mass of the trajectory. The shape of individual trajectories can be described by two properties: the gyradius of each location deviating from $(\bar{x}, \bar{y})$ and the direction of the trajectory along which most of the recorded locations occur, known as the principal axis. We followed the methods illustrated by Frank et al. (2013). First, a two-dimensional matrix known as the tensor of inertia was calculated for each active Twitter user's trajectory. The principal axis corresponds to the eigenvector with the largest eigenvalue. Considering the inherent anisotropy of the gyradius and direction across different individual trajectories, all individual trajectories were aligned in a common intrinsic reference frame. To do so, the principal axis of each trajectory was rotated pointing due west, and locations in each trajectory were translated anchored to $(\bar{x}, \bar{y})$ as $(x_i - \bar{x}, y_i - \bar{y})$ to ensure that the shape of each individual's trajectory fell in the reference frame. The intrinsic reference frame can be interpreted as a probability density function $P(x, y)$ of observing an individual. Instead of using the actual distance of the gyradius (Jurdak et al. 2015), which varies significantly among different individuals, we calculated the standard deviation $\sigma_x$ and $\sigma_y$ for each trajectory and normalized values by dividing their x- and y-coordinates as $x/\sigma_x$ and $y/\sigma_y$, respectively.

Figure 4 plots the heat map of the probability density function of the normalized locations of all active Twitter users. The shapes of the two heat maps are similar between the two cities. Interestingly, the observed shape is different from the teardrop shape reported by Frank et al. (2013), and the shape with higher values exists only on the positive side of the x-axis reported by Jurdak et al. (2015). The approaches of both studies were applied to individual trajectories of Twitter users at the national level. In contrast, the shape of the probability density function is similar to the ones reported by González, Hidalgo, and Barabási (2008), which were from the trajectories of mobile phone users at the city level. The probability density function demonstrates that people travel predominantly along their principal axis, with deviations in the orthogonal direction becoming shorter and less frequent as they move farther away from the origin. It shows that individuals spend the vast majority of their time in a few locations; for example, the locations in dark red regions are referred to as home locales, and the ones in the red regions are referred to as work locales. The shape of active Twitter users' trajectories confirms the preferential return behaviors that people (i.e., active Twitter users) frequently return to those locations, and therefore there is a chance that these transitions are recurrent across the group.

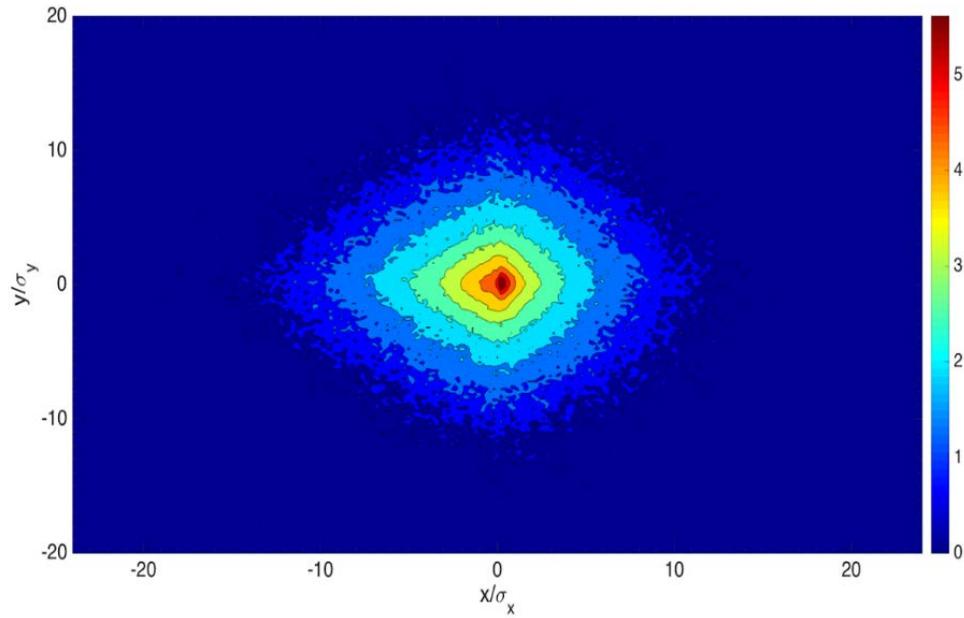

(A) Chicago

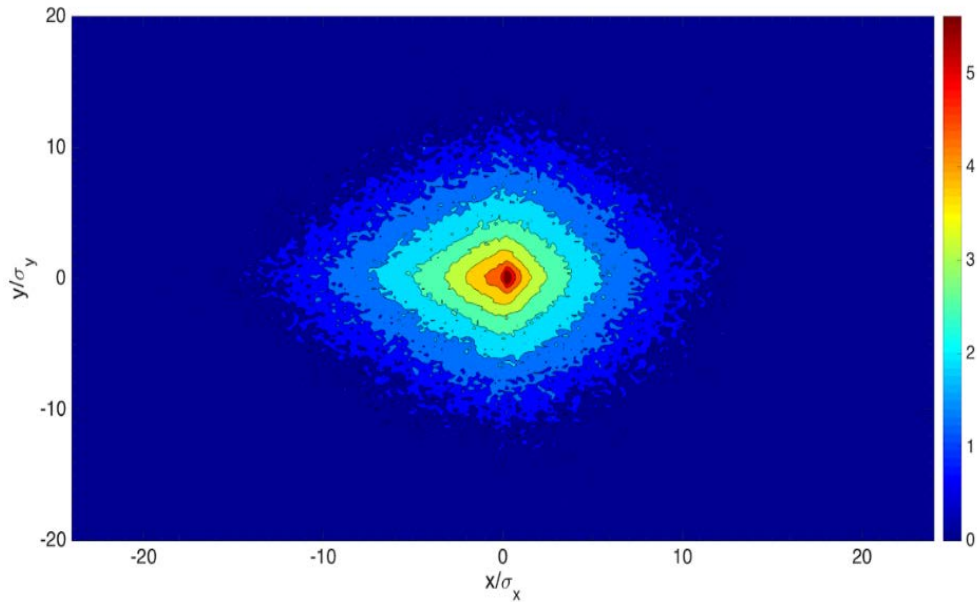

(B) Greater Boston

**Figure 4:** Probability density function $P(x, y)$ of observing an individual in the intrinsic reference frame. The origin corresponds to each individual's expected location, $x/\sigma_x$ and $y/\sigma_y$ represent the standard deviations of $(x, y)$ away from the origin in each respective axis, and $\sigma_y = 0$ corresponds to their principal axis rotated to the intrinsic reference frame. The maps are generated for active Twitter users in (A) Chicago and (B) Greater Boston.

## Daily location-based motifs

In this study, we investigated two types of motifs, i.e., LBMs and ABMs, by constructing mobility networks of the daily location histories of individuals. Given that previous studies have shown that people's weekday and weekend daily activities can be quite different (Schneider, Belik, et al. 2013; Jiang, Ferreira, and González 2017; Soliman et al. 2017), our study focuses on weekday activity motifs. There are 104,861 weekday equivalent observations of daily location histories from 10,001 active Twitter users in Chicago and 144,790 from 12,499 active Twitter users in Greater Boston. By constructing a mobility network for every daily location history, we could detect the LBMs using the VF2 algorithms as mentioned above.

Although a network motif often consists of only a few nodes, for a given network with N number nodes, a number of edge combinations exist ($2^{N^2-N}$). Because we are interested only in those mobility networks that reflect people's real daily transitional activities, we were able to significantly reduce the number of possible activity motifs by adopting the same two constraints used by Schneider, Belik, et al. (2013). The first constraint ensured that a user's daily location history starts and ends at the same activity location (in this case, a user's home location). The second constraint ensured each node in a daily mobility network (with $N > 1$) was visited at least once, which means each node had at least one ingoing and one outgoing edge. Figure 5 illustrates the 20 most recurrent daily LBMs in Chicago. The order of those motifs is ranked by their frequency. Because the network isomorphism does not support one-node networks, the one-node motif was calculated separately and was not plotted in the figure.

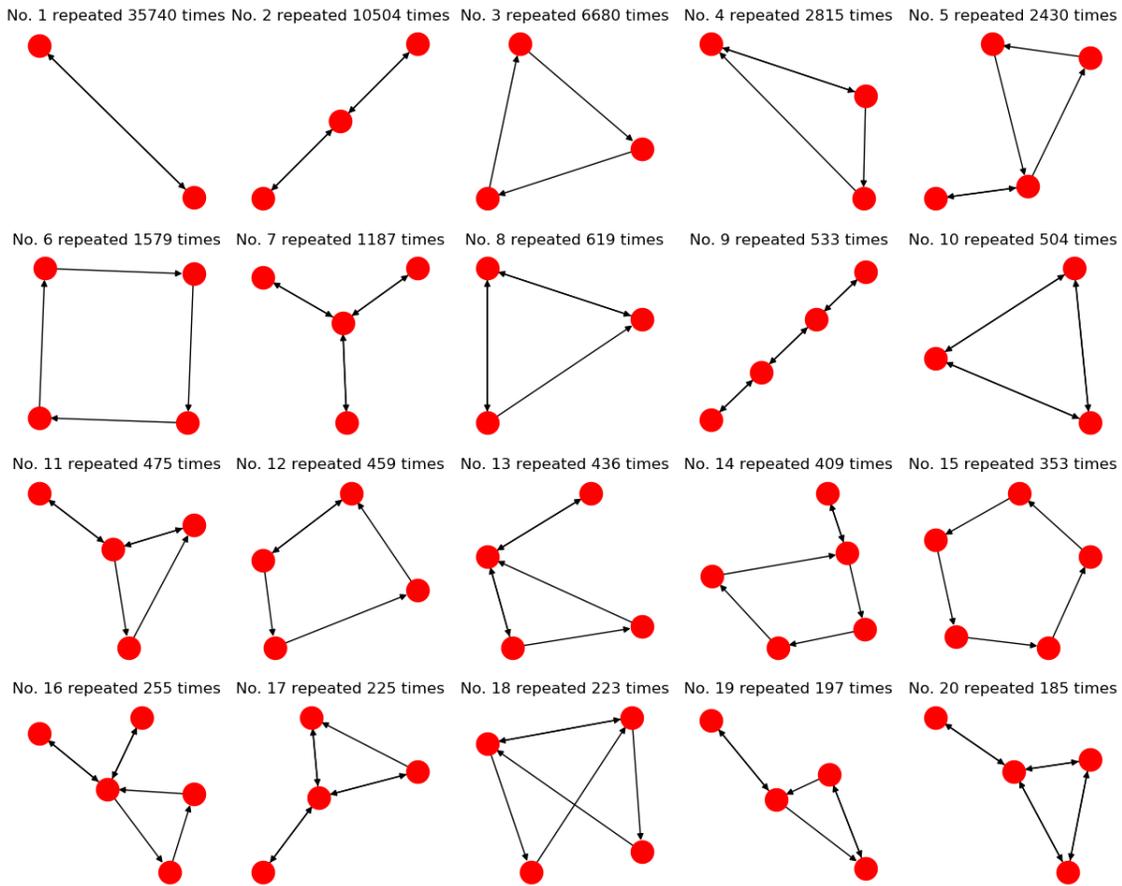

**Figure 5:** Top 20 most recurrent daily LBMs in Chicago; the order is ranked by frequency (note that the one-node motif is calculated separately and is not plotted in the figure)

Schneider, Belik, et al. (2013) identified 17 types of LBM using mobile phone data (in Paris) and travel survey data (in Chicago and Paris), which are shown in Figure 6A. A similar study using mobile phone data and travel survey data in Singapore showed that some of the LBMs were not significant, where 11 (out of 17) LBMs were identified (Jiang, Ferreira, and González 2017). In our study, 16 types of LBM are significant and consistent between Chicago and Greater Boston, which are shown in Figure 6B. In fact, 14 (out of 16) LBMs are identical to the ones identified by Schneider, Belik, et al. (2013). The LBMs with ID numbers 13, 16, and 17 are not significant in our study, but two new types of LBMs (one three-node LBM with ID 18 and one four-node LBM with ID 19) emerge as significant. Nevertheless, the 16 types of LBM account for over 83% of the total mobility networks in the two cities (83% in Chicago and 85% in Greater Boston). It should be noted that although the spatial accuracy of the locations in the three mobility data sources varies, the general LBMs are strikingly similar. The results suggest that (1) people's daily urban movements in visiting those frequently visited locations are significantly more recurrent than visiting other locations (2) the observations of LBMs are still valid when the movements are examined at the land use parcel level.

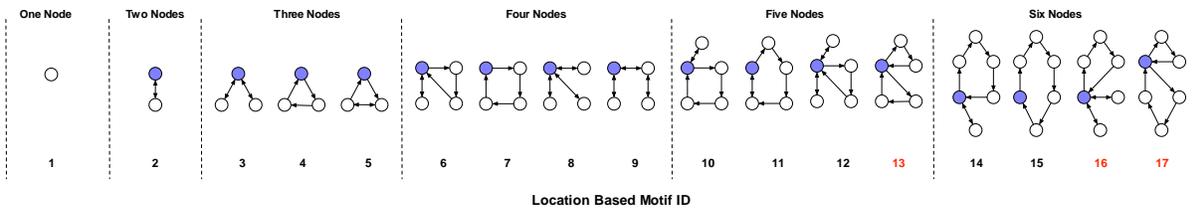

(A)

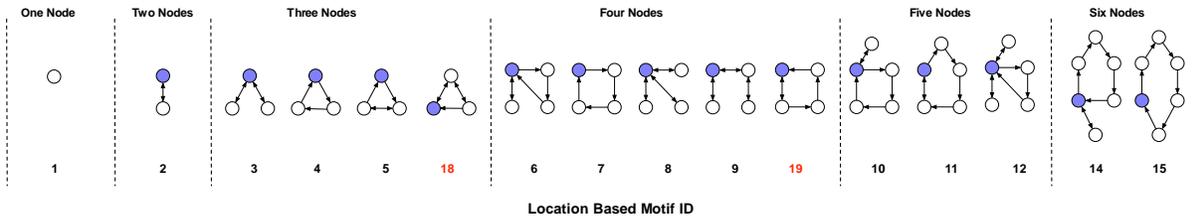

(B)

**Figure 6:** Identified LBMs in the daily mobility networks. (A) 17 types of LBM identified in Schneider, Belik, et al. (2013); (B) 16 types of LBMs in Chicago and Greater Boston identified in this study (the blue circle indicates the starting/finishing node in a user's daily mobility network).

Given the similarity between the LBMs identified in this study using geo-located Twitter data and the ones from using mobile phone and travel survey data, we compared the frequency distribution of those motifs together. Because of the differences in the LBMs identified in two studies (14 out of 19 were identical), instead of comparing the motifs one by one, we grouped them based on their node size, with a range from one to six. Figure 7 shows the percentage of the daily LBMs grouped by node size from using geo-located Twitter data (red, Chicago; green, Greater Boston), the surveys (dark red, Paris; blue, Chicago), and the phone data (light green, Paris). Even though those motifs came from multiple data sources and from different cities, the order and the percentage of a specific LBM group showed similar behavior, which suggests certain "unity" in people's daily mobility behavior.

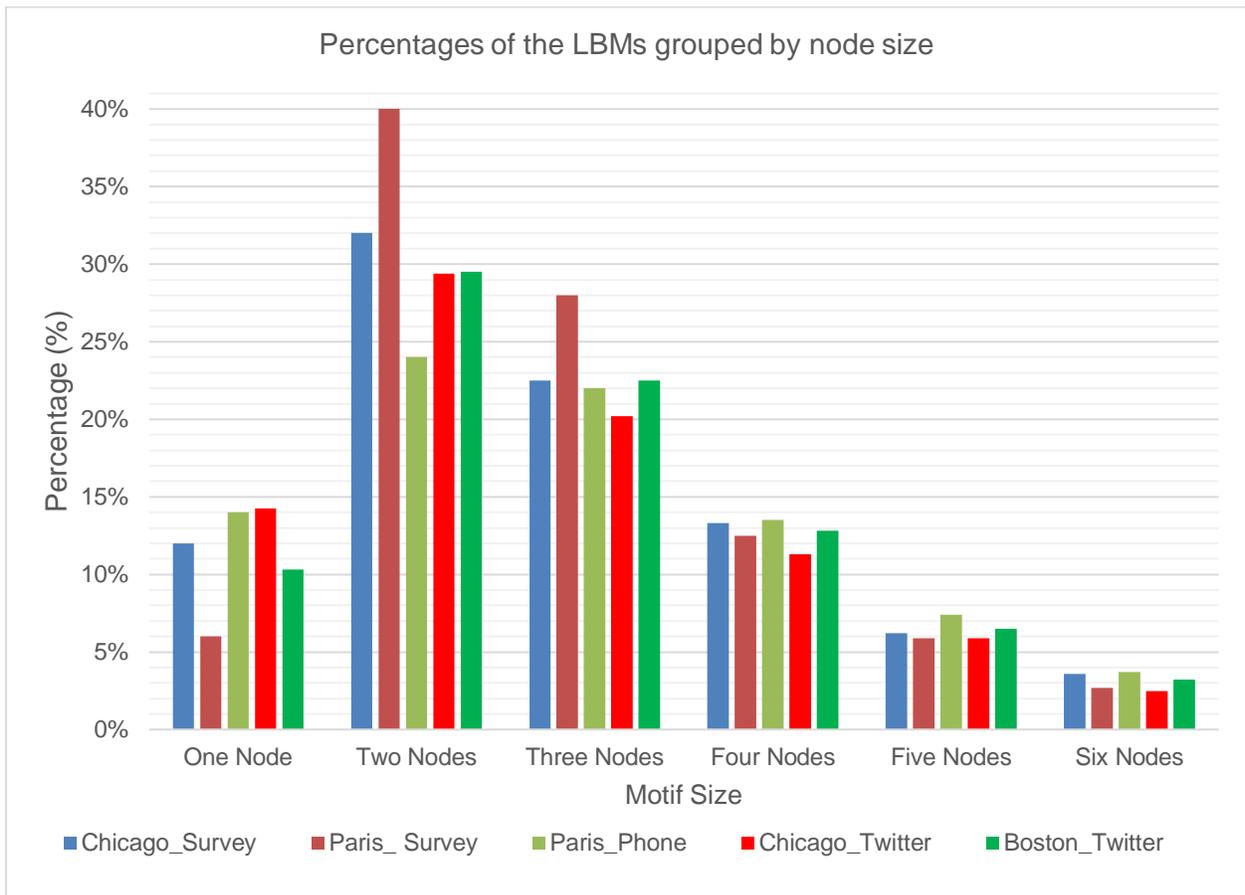

**Figure 7:** Percentage of daily LBMs grouped by node size from geo-located Twitter data (red, Chicago; green, Greater Boston) from our analysis, the surveys (dark red, Paris; blue, Chicago) from Schneider, Belik, et al. (2013), and the phone data (light green, Paris) from Schneider, Belik, et al. (2013).

The percentages of daily LBMs grouped by node size in Figure 7 appear to be similar among different cities, except the ones measured by survey data in Paris. However, the lack of accurate geographic locations in mobile phone and survey data makes it difficult to measure the variations in the spatial distances associated with people's daily LBMs. As the geographic locations captured in geo-located Twitter data are with high spatial accuracy, we calculated the spatial distances associated with various daily LBMs. First, since each user's home location was identified, we calculated the daily gyradius of each visited location (i.e., land use parcel) deviating from the home location, instead of deviating the center of mass of a daily trajectory. The results suggest that the average daily gyradius from home is 2.02 km in Chicago and 3.29 km in Greater Boston. Further, we calculated the average daily trip distance ($\widehat{d_l}$) and the average total daily distance ($\widehat{D_l}$) associated with different LBMs grouped by node size (Table 2). The results show that (1) $\widehat{d_l}$ ranges from 0.63 km to 1.11 km in Chicago and from 1.07 km to 1.86 km in Greater Boston (2) the LBMs with more nodes accumulate a larger $\widehat{D_l}$, and (3) Greater Boston has a larger $\widehat{D_l}$ than Chicago in all scenarios. Although the area of Greater Boston is nearly 8 times the size of Chicago, it is not proportional to the differences in the spatial distances. The results suggest that (1) most daily LBMs tend to concentrate at the small locale and (2) the spatial distances associate with the same type of LBM vary in different urban settings.

**Table 2**: The average trip distances and total daily distances associated with LBMs. Note that the distances are measured in km, and the percentage values correspond to the ones in Figure 7.

|  |  | Two Nodes | Three Nodes | Four Nodes | Five Nodes | Six Nodes | Seven Nodes+ |
|---|---|---|---|---|---|---|---|
| Chicago (548 km²) | Percentage | 29% | 20% | 11% | 6% | 3% | 16% |
|  | $\widehat{d_l}$ | 0.63 | 0.91 | 1.02 | 1.11 | 1.07 | 1.06 |
|  | $\widehat{D_l}$ | 6.42 | 10.04 | 12.52 | 15.13 | 16.89 | 22.7 |
| Greater Boston (4375 km²) | Percentage | 29% | 23% | 13% | 7% | 3% | 15% |
|  | $\widehat{d_l}$ | 1.07 | 1.56 | 1.82 | 1.86 | 1.75 | 1.78 |
|  | $\widehat{D_l}$ | 8.73 | 13.99 | 18.87 | 22.03 | 23.98 | 33.86 |

## Daily activity-based motifs

While the identified LBMs show promise in studying urban activity patterns, the exact details of the LBMs remain unclear. For example, the two-node LBM cannot test the assumption used in many prior studies that it is most likely the transition between home and work locations. With the ability to annotate the node types in people's daily mobility networks, we can now dissect the general LBMs into more detailed ABMs. For example, the top two-node ABMs in Chicago and Greater Boston are illustrated in Figure 8A and Figure 8B. Not surprisingly, the "Home-Work (noted as office)" motif was still the most prominent activity motif type. However, it accounted for only around 48% of the two-node mobility networks in both Chicago and Grater Boston. Some other two-node LBMs were considered significant in people's daily activities, such as "Home-Urban Mix", "Home-School", "Home-College", and others.

However, a deeper examination into the multi-node ($N > 2$) activity networks revealed more diverse transitional activity patterns. A noticeable pattern is that some nodes within a multi-node LBM were of the same activity type. One explanation is that a person may have multiple work locations, visit multiple friends (at different residential addresses), or go to multiple shopping locations. The activity types of the nodes in the identified LBMs show the unique and diverse daily activities of individuals. From the perspective of seeking activity patterns in the form of ABMs, we can merge the intermediate nodes with the same activity type but still preserve the order of transitions in people's daily activities. By applying this procedure, we identified 17 types of ABMs in Chicago and 21 in Greater Boston. Figure 8 illustrates the most recurrent ABMs in the two cities (two one-node ABMs were calculated separately).

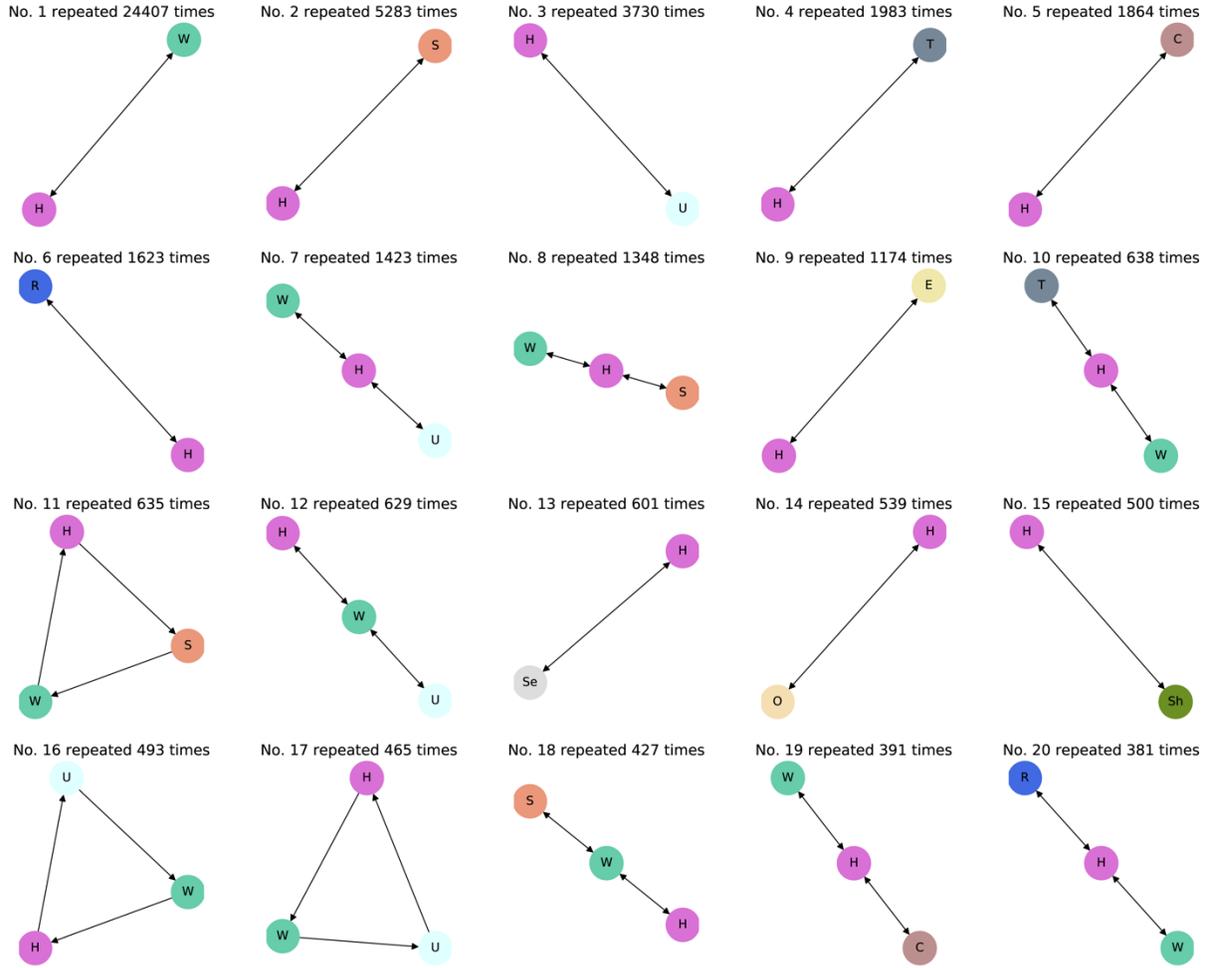

(A) Chicago

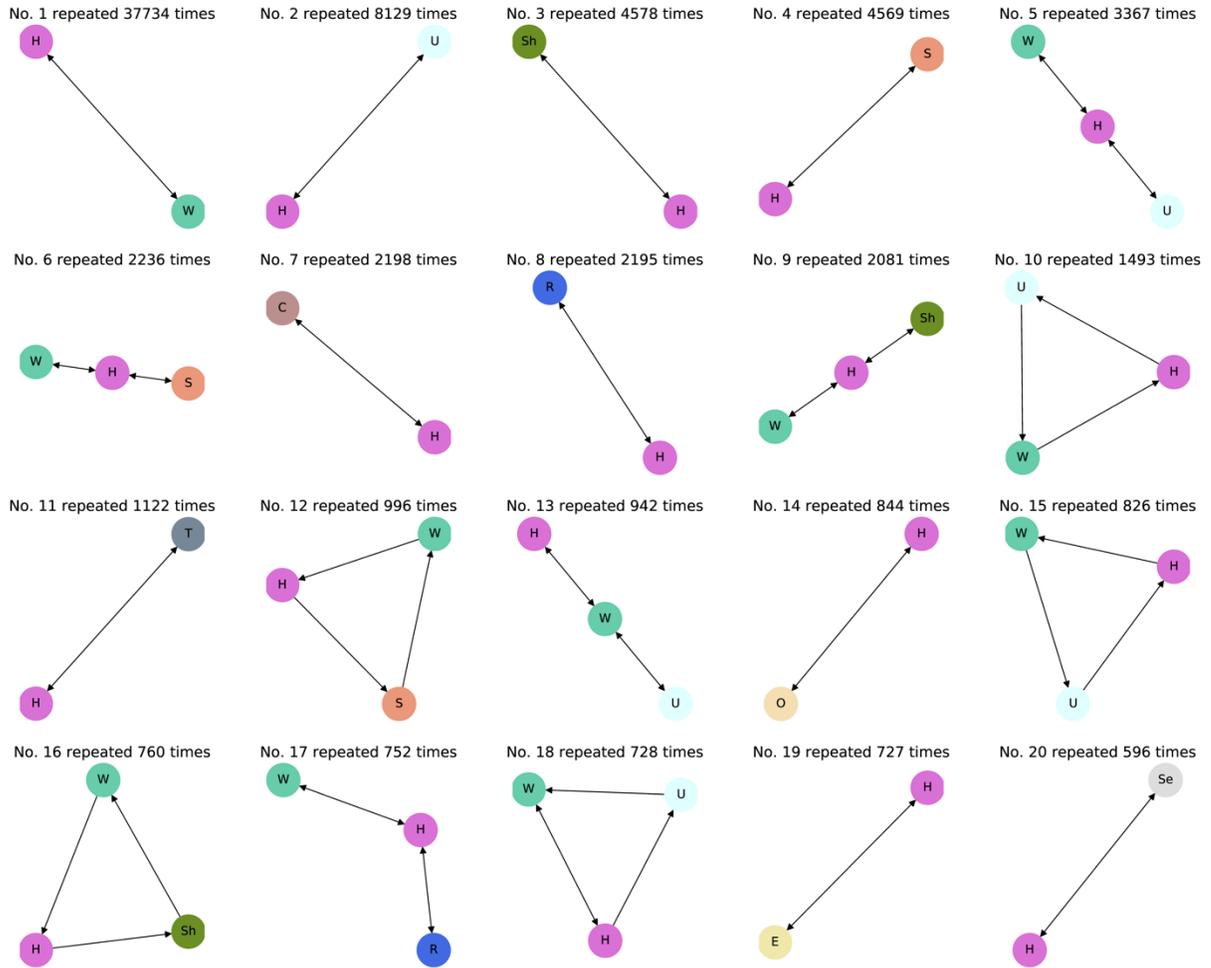

(B) Greater Boston

**Figure 8:** Most recurrent daily ABMs in (A) Chicago and (B) Greater Boston. The order is ranked by frequency (the one-node motif was calculated separately and is not plotted in the figure). Home (H): orchid; Work/Office (W): medium aquamarine; School (S): dark salmon; Urban mix (U): light cyan; Transportation (T): light slate gray; College (C): rosy brown; Residential (R): royal blue; Shopping (Sh): olive drab; Entertainment (E): pale goldenrod; Civic service (Se): gray; Other (O): wheat

The percentages of the common daily ABMs in Chicago and Greater Boston are summarized in Figure 9. Most of the ABMs have similar behaviors between the two cities, except for a relatively larger difference in motif 11 (i.e., a two-node ABM representing home-shopping activity). Nevertheless, these common ABMs describe over 57% (57.5% in Chicago and 58.3% in Greater Boston) of the transitional behaviors in daily activities in the two cities.

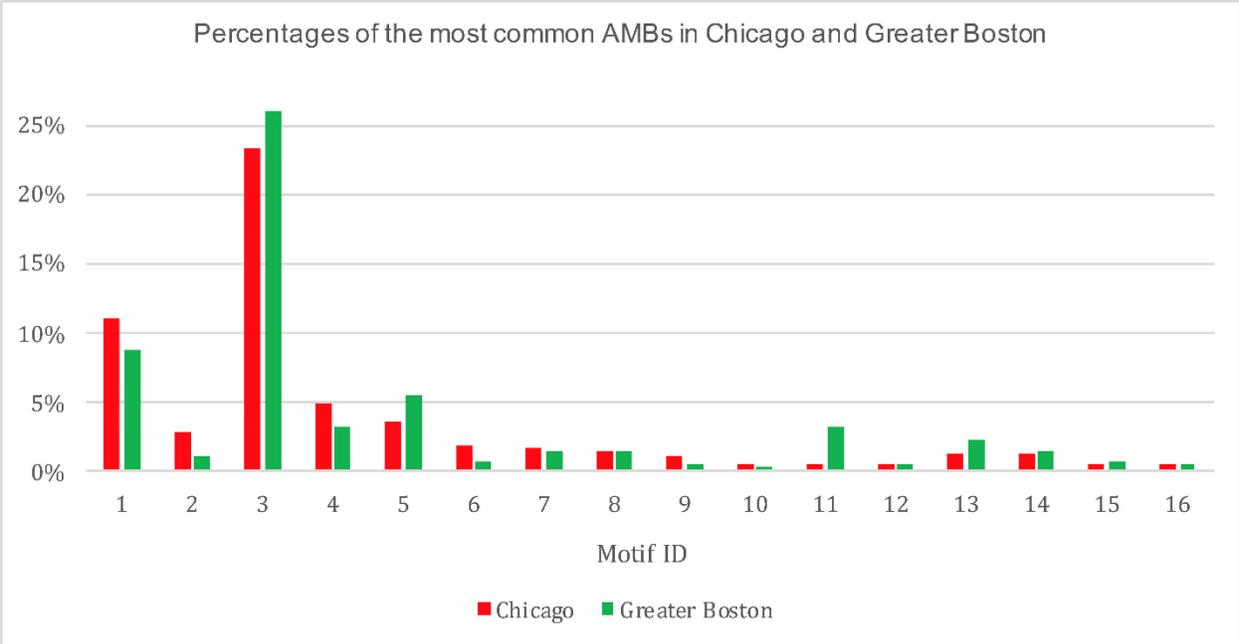

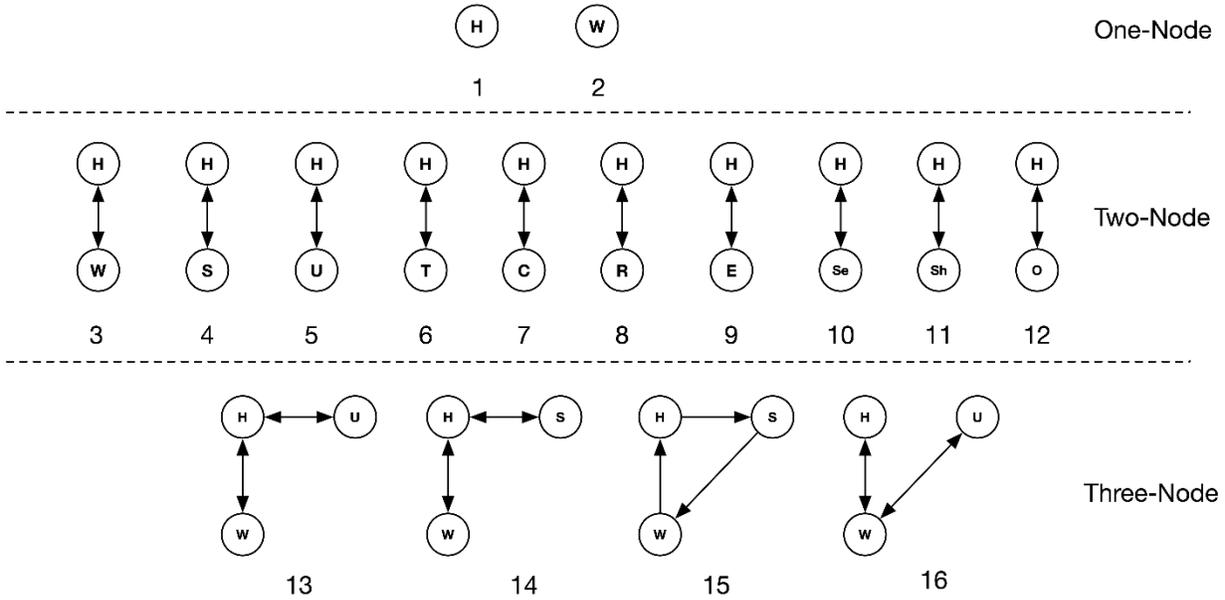

**Figure 9**: Percentages of common daily LBMs in Chicago and Greater Boston. H: home; W: work/office; S: school; U: urban mix; T: transportation; C: college; R: residential; E: entertainment; Se: civic service; Sh: shopping; O: others

Another major advantage of using geographic context-aware Twitter data to study activity patterns is that we can measure accurate spatial distances in the transition from not only one location to another (LBMs), but also from one type of activity to another (ABMs). For example, we calculated the average trip distances ($\widehat{d_a}$) in several two-node LBMs (Table 3). While $\widehat{d_a}$ between home and other activities are slightly shorter in Chicago than in Greater Boston, an interesting observation is that $\widehat{d_a}$ between home and school (H-S) is 4.52 km in Chicago and 3.39 in Greater Boston. It suggests that on average people in Chicago take longer daily trips from home to school than people do in Greater Boston. Although the result should be verified by some official data or to be confirmed with future studies in other cities, it does provide a link to the spatial homogeneity of the distribution of urban infrastructures/land use (schools in this context) can impact the spatial distances in people's daily LBMs in different urban settings. Further, we calculated the average total daily distance ($\widehat{D_a}$) associated with different ABMs grouped by node size. The results report a similar trend observed in the spatial distances associated with LBMs (1) the ABMs with more nodes accumulate a larger $\widehat{D_l}$, and (2) Greater Boston has a larger $\widehat{D_a}$ than Chicago in all scenarios.

**Table 3**: The average trip distances and total daily distances associated with ABMs. Note that the distances are measured in km, and the percentage values correspond to the ones in Figure 9.

|  | H-W | H-S | H-Sh | H-T | Two-Node | Three-Node | Four-Node | Five-Node | Six-Node | Seven-Node+ |
|---|---|---|---|---|---|---|---|---|---|---|
| Percentage |  |  |  |  | 47% | 26% | 9% | 3% | 1% | <1% |
| Chicago | 7.32 | 4.52 | 5.00 | 4.32 | 4.48 | 11.26 | 16.20 | 20.68 | 27.29 | 31.06 |
| Percentage |  |  |  |  | 46% | 27% | 8% | 3% | 1% | <1% |
| Boston | 9.15 | 3.39 | 6.65 | 7.00 | 7.63 | 17.37 | 24.99 | 33.51 | 48.00 | 61.01 |

H: home; W: work/office; S: school; Sh: shopping; T: transportation

# 5. Conclusions and Discussions

The overarching goal of this study is to gain insights into people's daily urban activity patterns. People's daily activities are complex and vary by individuals. Existing studies using mobile phone call data to infer people's daily activities have suggested that, collectively, people's movements across the urban space can be well modeled, are highly predictable with a tendency for people to return to previously visited locations, and most of the transitional behaviors in traversing from one activity location to another can be described by a set of distinct and recurrent mobility motifs. In this connection, the fundamental hypothesis of this study is that there are significant regularities in the transitions among people's daily activities (i.e., activity patterns) despite the large variation in human activities (e.g., different individuals, different cities, and different dates).

The novelty of this study is the synthesis of geo-located tweets with detailed land use parcel maps to enhance the geographic context of the mobility dataset for inferring specific activities (e.g., home, work, and shopping activity). To ensure those activities were actively observed, we utilized a criterion to select active users for further analysis. A mobility network was constructed for every active user's daily location history. For LBMs, the nodes in a network are visited activity locations, and the edges are movements transitioning from one location to another. For ABMs, the nodes are semantically labeled activity types, whereas the edges represent transitional behaviors traversing those activities. By analyzing the shape of the collective location histories of the active users, we identified preferential return patterns that confirm the existence of recurrent transitional behaviors among different urban activities. Further exploration into the isomorphic structure within the collection of daily mobility networks uncovered 16 types of unique LBMs, which described over 83% of the daily mobility networks in the two cities and were comparable to the ones from previous studies using mobile phone data and travel surveys. With the detailed and semantically labeled transitions between each two activity types, we were able to further dissect the general location-based motifs into 16 common ABMs that described over 57% (57.5% in Chicago and 58.3% in Greater Boston) of the transitional behaviors in the daily activities in the two cities. Note that although the general activity motif patterns of both LBMs and ABMs in Chicago and Greater Boston were similar, the spatial heterogeneity of urban land use can potentially contribute to the variations of spatial distances associated with different activity motifs grouped by node size. This study demonstrates that, with geographic context-aware Twitter data, we can gain further insights into revealing the unique activity motifs that form the fundamental elements embedded in complex urban activities.

The ability to track people's movement is critical for studying urban activity patterns. While mobile phone data and travel surveys are two popular data sources, we discussed the advantages and disadvantages of both data sources, focusing on data accessibility and spatiotemporal accuracy/granularity. For example, because of the low spatial accuracy of mobile phone data, it is difficult to engineer such a dataset for inferring people's activities other than staying home or working, which is why prior studies have focused on primary mobility patterns (e.g., LBMs), where people's visited locations can be inferred but the semantics/types of those locations are unknown). In terms of data accessibility and availability, the scalability of these datasets for different cities/regions is also challenging. Therefore, the combination of publicly accessible geo-located Twitter data and parcel-level land use maps provides a unique opportunity to examine not only LBMs but also to examine detailed ABMs.

It should be noted that there are limitations in using geo-located tweets to track people's whereabouts and explore their activity patterns. First, although this study showed that Twitter user populations correlate well with the actual population at the census tract level in the two cities, the Twitter user population may not be a representative sample of the actual population with its particular demographic composition (Yin, Chi, and Van Hook 2018) and skewed popularity among younger people (Greenwood, Perrin, and Duggan 2016). Indeed, the prominent existence of the two-node ABMs, such as "Home-School" and "Home-College", indicates this issue. Some studies have attempted to develop methods to make Twitter data more representative (Zagheni and Weber 2015), and others have explored the different demographic characteristics of human mobility patterns with geo-located Twitter data (Luo et al. 2016). However, until the representativeness of Twitter data can be addressed, identified activity patterns will likely not be fully generalizable to the whole population.

Second, the methodologies developed in this study utilized geo-located Twitter data to track the movements of individuals. Ideally, the methods would work better for movement data capable of tracking people's locations continuously, such as using GPS units. Both mobile phone call data and geo-located tweets are treated as proxies for tracking user locations. Owing to variations in people's behaviors when using mobile phones or Twitter, the latent locations might not be recorded—hence, those activities were not considered. For example, suppose a user posts one tweet at home in the morning and another at work at noon; the whereabouts or activities of this user during the gap between tweets were unknown. Related to this, the temporal characteristics of the activity motifs in Chicago and Greater Boston were not explored in this study, which should be addressed and compared in future studies. Although imposing a strict criterion to select active users helps minimize the impact of missing latent locations, the included population in the studies may skew toward people who are actively engaging in particular activities, such as making phone calls or tweeting. Therefore, methods that address such a situation should be employed in future studies.

Finally, the synesis of geo-located Twitter data with parcel-level land use maps poses potential privacy concerns. On one hand, the high spatial accuracy enables us to examine detailed spatial and temporal activity patterns. On the other hand, those derived activity locations can expose where people live, work, and study, etc., which can raise significant privacy concerns of individuals. In this study, we have taken some precautions in processing the data, such as removing the Twitter user identities, reassigning IDs to the land use parcel, and using only the types of land use parcels in the analysis. However, more widely accepted guidelines and practices should be developed for future studies involving observation of individuals with high spatial accuracy and temporal granularity.

# Acknowledgments

We thank the anonymous reviewers for their constructive comments on earlier versions of the paper. This research was supported in part by the National Science Foundation (Awards #1541136 and #1823633); the Eunice Kennedy Shriver National Institute of Child Health and Human Development (Award #P2C HD041025); and the Social Science Research Institute, Population Research Institute, and Institute for Computational and Data Sciences of the Pennsylvania State University.

JUNJUN YIN is an Assistant Research Professor in the Computational and Spatial Analysis Core at The Pennsylvania State University, University Park, PA 16802. E-mail: jyin@psu.edu. His research interests include computational geography approaches and geospatial Big Data to model human-urban environment interactions about urban mobility, accessibility, and sustainability.

GUANGQING CHI is Professor of Rural Sociology, Demography, and Public Health Sciences in the Department of Agricultural Economics, Sociology, and Education and Director of the Computational and Spatial Analysis Core at The Pennsylvania State University. E-mail: gchi@psu.edu. His research interests focus on the socio-environmental systems, aiming to understand the interactions between human populations and built and natural environments and to identify important assets to help vulnerable populations adapt and become resilient to environmental changes by developing and implementing spatial and big data analytic methods.